\documentclass[a4paper,11pt]{amsart}
\begin{document}

\title{{\bf  GRAVITATIONAL WAVES AND ETHER'S WIND}}
\author{Angelo Loinger}
\date{}
\address{Dipartimento di Fisica, Universit\`a di Milano, Via
Celoria, 16 - 20133 Milano (Italy)}
\email{angelo.loinger@mi.infn.it}
\thanks{Submitted to \emph{Il Nuovo Saggiatore} -- Bollettino della Societ\`a Italiana di Fisica}

\begin{abstract}
A very recent research validates observationally the theoretical
demonstrations of the physical non-existence of the gravitational
waves.
\end{abstract}

\maketitle

\vskip1.20cm
Recently (August 14th, 2003) it has been published on \emph{Los
Alamos Archive} (arXiv:gr-qc/0308050 v1) an observational paper by
368 co-authors (B.Ab\-bott \ldots J.Zweizig) entitled ``Setting
upper limits on the strength of periodic gravitational waves using
the first science data from the GEO600 and LIGO detectors''. \par
Its abstract is as follows: ``Data collected by the GEO600 and
LIGO interferometric gravitational wave detectors during their
first observational science run were searched for continuous
gravitational waves from the pulsar J1939+2134 at twice its
rotation frequency. Two independent analysis methods were used and
are  demonstrated in this paper: a frequency domain method and a
time domain method. Both achieve consistent null results, placing
new upper limits on the strength of the pulsar's gravitational
wave emission. A model emission mechanism is used to interpret the
limits as a constraint on the pulsar's equatorial ellipticity.''
\par
 \emph{These null results could have been foreseen} -- and in a very simple way.
 A fundamental article of 1917 by T.Levi Civita
 \cite{1} has demonstrated that the gra\-vi\-ta\-tio\-nal waves are mere
 \emph{formal} undulations, fully destitute of energy and
 momentum, \emph{only} endowed with a \emph{false} (pseudo) tensor
 (i.e. a \emph{non}-tensor) of energy-momentum. As a matter of
 fact, Einstein had always serious doubts about the
 \emph{physical} reality of the gravitational waves, see e.g. his
 paper with Rosen of 1935 \cite{2}, even if the notion ``gravitational wave''
 had been theorized by him in 1916 \cite{3}. But in this paper he
 investigated the approximate \emph{\textbf{linearized}} version of the exact
 general relativity (version whose substrate is simply Minkowski
 spacetime), which resembles the Minkowskian formulation of e.m.
 Maxwell theory, and which has an invariant character only under
 the transformations of Lorentz group. On the contrary, Levi-Civita
 \cite{1} made a frontal attack to the analogous problem of the
 \emph{exact} GR, for which the variable metric \emph{\textbf{is}}, in
 essence, spacetime -- and \emph{not} a conventional field
 propagated through a fixed substrate. \par
 In recent years I have given specific and stringent proofs of the
 physical non-existence of the gravitational waves \cite{4}, in
 particular of the absence of any generation mechanism whatever.
 \par
 It is a pity that in 2003 the above 368 physicists base
 essentially their belief in the real existence of gravitational
 waves on the cited Einstein's paper of 1916 \cite{3}, fully
 neglecting some basic concepts of the \emph{exact} GR \cite{5}.
 Of course, they have the possibility of persevering on their road
 with the aim to lower more and more the upper limits of the
 strength of the gravitational waves. And in fact they write that
 ``further improvements are planned'': a vain chase to nothing.
 In the Forties of the past century a distinguished
 experimentalist, Quirino Majorana (uncle of Ettore M.), had
 succeeded in lowering to the intensity of a very gentle breeze
 the upper limit of the
 strength of terrestrial ether's wind; he used just a Michelson
 interferometer. \par
 An Italian proverb says: ``Chi si contenta gode''.

\end{document}